# Analysis of Daily Streamflow Complexity by Kolmogorov Measures and Lyapunov Exponent

**Authors**

Dragutin T. Mihailović[a], Emilija Nikolić-Đorić[a], Ilija Arsenić[a], Slavica Malinović-Milićević[b], Vijay P. Singh[c], Tatijana Stošić[d], Borko Stošić[d*],

**Affiliations**

[a]Faculty of Agriculture, University of Novi Sad, Dositej Obradovic Sq. 8, 21000 Novi Sad, Serbia; guto@polj.uns.ac.rs (D.T.M); emily@polj.uns.ac.rs (E. N.-Đ.); ilija@polj.uns.ac.rs (I.A.)

[b]ACIMSI-Center for Meteorology and Environmental Modelling, University of Novi Sad, Dositej Obradovic Sq. 7, 21000 Novi Sad, Serbia; slawica@sbb.rs (S.M.M.)

[c]Department of Biological and Agricultural Engineering and Zachry Department of Civil Engineering, Texas A and M University, College Station, TX 77843-2117, USA; vsingh@tamu.edu  (V.P.S)

[d]Departamento de Estatística e Informática, Universidade Federal Rural de Pernambuco,  Rua Dom Manoel de Medeiros s/n, Dois Irmãos, 52171-900, Recife, Brazil; tastosic@gmail.com (T.S.); borkostosic@gmail.com (B.S.)

**Abstract**

Analysis of daily streamflow variability in space and time is important for water resources planning, development, and management. The natural variability of streamflow is being complicated by anthropogenic influences and climate change, which may introduce additional complexity into the phenomenological records. To address this question for daily discharge data recorded during the period 1989–2016 at twelve gauging stations on Brazos River in Texas (USA), we use a set of novel quantitative tools: Kolmogorov complexity (KC) with its derivative associated measures to assess complexity, and Lyapunov time (LT) to assess predictability. We find that all daily discharge series exhibit long memory with an increasing downflow tendency, while the randomness of the series at individual sites cannot be definitively concluded. All Kolmogorov complexity measures have relatively small values with the exception of the USGS (United States Geological Survey) 08088610 station at Graford, Texas, which exhibits the highest values of these complexity measures. This finding may be attributed to the elevated effect of human activities at Graford, and proportionally lesser effect at other stations. In addition, complexity tends to decrease downflow, meaning that larger catchments are generally less influenced by anthropogenic activity. The correction on randomness of Lyapunov time (quantifying predictability) is found to be inversely proportional to the Kolmogorov complexity, which strengthens our conclusion regarding the effect of anthropogenic activities, considering that KC and LT are distinct measures, based on rather different techniques.

---

[*] Corresponding author tel. +55 81 996593064
   E-mail address: borkostosic@gmail.com (B. Stosic)

**1. Introduction**

Streamflow varies in space and time and investigation of its variability is essential for planning, operation, and management of water resources systems. It is significantly influenced by human activities and climatic change, and represents a complex system. In the past two decades, several information measures and methods of nonlinear dynamics have been employed to explain the complex and chaotic nature of streamflow [1-9]. One of the information measures for quantifying randomness of streamflow is the Kolmogorov complexity (KC). This measure and its derivatives - Kolmogorov spectrum, Kolmogorov complexity spectrum highest value (KCM) [10], running Kolmogorov complexity (RKC), and mean Kolmogorov complexity (MKC) - provide information complementary to that of entropy, offering additional insight into the behavior of complex systems. However, Kolmogorov complexity (KC) has rarely been used to analyze streamflow time series, although it is capable of providing supplementary description (to that of entropy) on the complexity or degree of randomness in highly complex systems. Some attempts to use KC and its derivatives for analyzing streamflow were made by [6,8,10]. The KC and its derivatives are important to use, because randomness is a key characteristic that has important implications for streamflow modeling and prediction [11,12]. Another complementary measure important for streamflow analysis is the largest Lyapunov Exponent (LE), because it can detect the presence of deterministic chaos in complex hydrological systems and quantify the predictability of potential future outcomes [13-15].

The objective of this paper therefore is to investigate the complexity, chaotic behavior, and predictability of daily streamflow of Brazos River in Texas (USA) for the period 1989–2016, using KC, KC spectrum, KCM, MKC, RKC, and LE. These analyses reveal those aspects of the phenomenon that are complementary to those of the classical statistical approach. More precisely, the empirical study performed in the current work (that should not be confused with an attempt to modeling), sheds new light on the general properties of the observed time series (intrinsic randomness and predictability). These aspects should in turn be taken into account when devising new heuristic or mathematical models, but that can also be useful for detecting "unusual" behavior at some selected locations, such as at the USGS station 08088610 (Brazos River near Graford, TX), where high KC complexity may be attributed to the nearby Morris Sheppard Hydroelectric Power plant.

The paper is organized as follows. Part 2 describes the Kolmogorov complexity and information measures derived from it (Kolmogorov complexity spectrum and its highest value, and running and mean Kolmogorov complexity), Lyapunov exponent, and Lyapunov time. Part 3 provides information on streamflow data and gauging locations. Finally, part 4 presents the results obtained, discusses the nonlinear feature of streamflow time series, Kolmogorov complexity measures and complexity spectrum, and Lyapunov exponent; and the predictability of daily discharge. The concluding remarks are given in part 5.

## 2. Methods

### 2.1. Kolmogorov Complexity and Its Derivatives

#### 2.1.1. Kolmogorov Complexity

The Kolmogorov complexity [16] stems from algorithmic information theory, which has been increasingly employed over the past 15 years for characterization of complex phenomena [6,8,10,17,18]. Let $X$ denote streamflow and $x$ its specific value. Kolmogorov complexity $K(x)$ of an object $x$ is the length, in bits, of the smallest program that can be run on a universal Turing machine $U$ and that prints object $x$ (a Turing machine is a mathematical model of an abstract machine, which can use a predefined set of rules to determine a result from a set of input variables). While the complexity $K(x)$ is not directly computable for an arbitrary object $x$, in practice it is approximated by the size of the ultimate compressed version of $x$ [19,20]. More precisely, a binary object $x$ is compressed, and the size of the compressed object is identified with Kolmogorov complexity $K(x)$. Lempel and Ziv [19] suggested an algorithm (LZA) for calculating KC of a time series. We shortly describe the calculation of the KC complexity of a time series $X(x_1, x_2, x_3, ..., x_N)$ by the LZA algorithm. It includes the following steps. (1) Encoding the time series by creating a sequence $S$ of the characters 0 and 1 written as $s(i), i = 1,2, ..., N$, according to the rule $s(i) = 0$ if $x_i < x_t$ or 1 if $x_i > x_t$, where $x_t$ is a threshold. The threshold is commonly selected as the mean value of the time series [17], while other encoding schemes are also available [21]; (2) Calculating the complexity counter $c(N)$. The $c(N)$ is defined as the minimum number of distinct patterns contained in a given character sequence. The complexity counter $c(N)$ is a function of the length of the sequence $N$. The value of $c(N)$ is approaching an ultimate value $c(N)$ as $N$ is approaching infinity, i.e. $c(N) = O(b(N))$ and $b(N) = N/log_2 N$; (3) Calculating the normalized information measure $C_k(N)$, which is defined as $C_k(N) = c(N)/b(N) = c(N)N/log_2 N$. For a nonlinear time series, $C_k(N)$ varies between 0 and 1, although it can be larger than 1 [18]. The LZA algorithm is described in condensed form [19,22]. Note that the pattern is a sequence in the coded time series which is unique and non-repeatable. A flow chart for calculation of KC of a streamflow series $X(x_1, x_2, x_3, ..., x_N)$ using the LZA algorithm is shown in Fig.1.

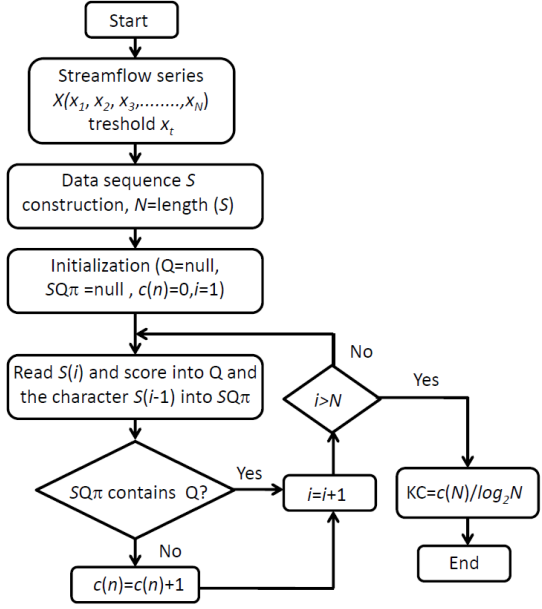

**Fig. 1.** Flow chart for calculation of the Kolmogorov complexity (KC) using the Lempel-Zev algorithm (LZA) [23]

.

### 2.1.2. Kolmogorov Complexity Spectrum and Its Highest Value

The Kolmogorov complexity of streamflow has two weaknesses: (i) It cannot distinguish between time series with different amplitude variations and that with similar random components; and (ii) in the conversion of a time series into a binary string, its complexity is unseen in the rules of the applied procedure. Therefore, in defining a threshold for a criterion for coding, some information about the composition of time series could be lost. In streamflow analysis, two measures were used: (i) Kolmogorov complexity spectrum (KC spectrum) and (ii) the highest value of KC spectrum, introduced by [10] who described the procedure for calculating the KC spectrum. Figure 2 schematically shows how to calculate the KC spectrum $C(c_1, c_2, c_3, \ldots, c_N)$ for streamflow series $X(x_1, x_2, x_3, \ldots, x_N)$. This spectrum allows investigating the range of amplitudes in a time series that represents a complex system with highly enhanced stochastic components. It may be noted that for a large number of samples of a time series, the computation of KC spectrum can be computationally challenging. Therefore, it is reasonable to divide the domain, including all values between the minimum ($c_{min}$) and maximum ($c_{max}$), in the time series into subintervals, which then are used as thresholds. In this study, 500 thresholds were used to obtain all spectra. The highest value $K_m^C$ as in this series, i.e., $K_m^C = max\{c_i\}$, is the highest value of Kolmogorov complexity spectrum (KCM), as seen in Fig. 2.

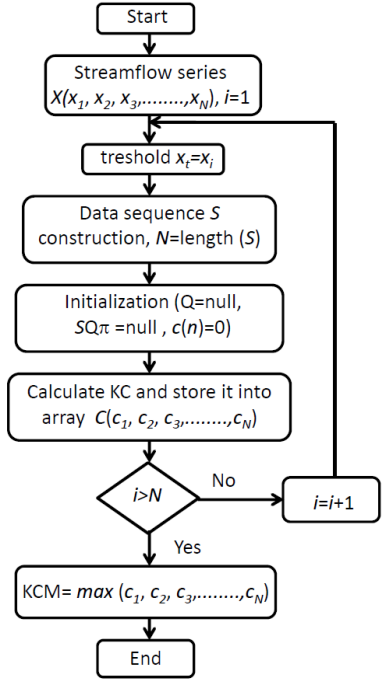

**Fig. 2.** Flow chart for calculation of the Kolmogorov complexity spectrum and its highest value (KCM) [23]

2.1.3. Mean Kolmogorov Complexity

Using the mean-value theorem and the Simpson's rule formula, the mean Kolmogorov complexity $K_C^M$ (MKC) can be calculated as

$$K_C^M = \frac{\Delta c}{3(c_n - c_0)}(c_0 + 4c_1 + 2c_2 + 4c_3 + 2c_4 + \ldots + 4c_{n-1} + 4c_n) \qquad (1),$$

here $c_0 = min\{c_i\}$ and $c_n = max\{c_i\}$, while the interval $[c_0, c_n]$ is divided into an even number $n$ of subintervals of equal length $\Delta c = (c_n - c_0)/n$ using the $n + 1$ points.

2.1.4. Running Kolmogorov Complexity

In statistical hydrology, a running mean [24] is often calculated to analyze data by creating series of averages of different subsets of the full data set that is a type of finite impulse response filter. This procedure was used to calculate the running Kolmogorov complexity (RKC). For a given series of streamflow values we extracted a fixed window (of size 1095 in this study) as in running mean procedure, and then the procedure for calculating the KC (Fig. 1) is applied. After that, the window is moved forward and the KC algorithm is applied, until the end of time streamflow series is reached.

2.2. Lyapunov Exponent and Lyapunov Time of Streamflow Time Series

The Lyapunov exponent of a dynamical system is a quantity that characterizes the rate of separation of infinitesimally close trajectories [25] in parameter space. Positive Lyapunov exponent (LE) indicates that small fluctuations can lead to drastically different system behavior (small differences in the initial state lead to large differences in a later state). The term "butterfly effect" was coined by Edward Lorenz to describe this effect in a metaphorical example where flapping of the wings of a butterfly may lead to a faraway tornado in a distant future.

Quantitatively, two trajectories in phase space with initial separation $\delta X$ diverge at a rate given by $|\delta X(t)| \approx e^{\lambda t}|\delta X(0)|$ where $\lambda$ is the Lyapunov exponent. It should be noted that the divergence is treated within the linear approximation. Because the rate of separation can be different for different orientations of the initial separation vector, there is a spectrum of Lyapunov exponents whose largest value is commonly the LE. A positive value of this exponent is usually taken as an indication that the system is chaotic. In this study, we obtained the LE for the standardized daily discharge time series by applying the Rosenstein algorithm [26], which was implemented in MATLAB program [27]. In this algorithm $\lambda$ is calculated as

$$\lambda = \lim_{\tau \to \infty} \lim_{\varepsilon \to 0} \frac{1}{\tau} ln\left(\frac{|x(\tau) - x_\varepsilon(\tau)|}{\varepsilon}\right), \qquad (2)$$

where $|x(0) - x_\varepsilon(0)| = \varepsilon$.

This algorithm is fast, easy to apply, and robust to changes in embedding dimension, reconstruction delay, length of time series and noise level. The applied MATLAB program calculates the proper embedding dimension and reconstruction delay. The value of embedding dimension is selected by FNN (False Nearest Neighbors) method or symplectic geometry method in the case of high noisy data [28,29]. The proper time delay is the lag before the first decline of autocorrelation value below $1/e$. For data with nonlinear dependency time delay is obtained applying mutual information criteria [30]. The Lyapunov exponent relates to the predictability of measured time series which includes deterministic chaos as an inherent component. Model predictability is here understood as the degree to which a correct prediction of a system's state can be made either qualitatively or quantitatively. Causal determinism has a strong relationship with predictability, in that perfect predictability requires strict determinism. However, lack of predictability does not necessarily mean lack of determinism, where limitations on predictability could be caused by factors such as a lack of information or high level of randomness. There exists no general rule to predict the time evolution of systems far from equilibrium, e.g. chaotic systems that do not approach an equilibrium state. Their predictability usually becomes progressively worse with time, and to quantify predictability, the rate of divergence of system trajectories in phase state can be measured (for example, Kolmogorov-Sinai entropy, and Lyapunov exponents [25]). In stochastic analysis a random process is considered predictable if it is possible to infer the next state from previous observations. In many models, however, randomness is a phenomenon which "spoils" predictability. Deterministic chaos does not mechanically denote total predictability but means that at least it improves the prognostic power. In contrast, stochastic trajectories cannot be projected into future. If $\lambda > 1$ then streamflow is not chaotic but is rather stochastic, and predictions cannot be based on chaos theory. However, if $0 < \lambda < 1$ it indicates an existence of chaos in streamflow. In that case, one can compute the approximate time (often called Lyapunov time) limit for which accurate prediction for a chaotic system is a function of LE. It designates a period, when a certain process (physical, mechanical, hydrological, quantum, or even biological) moves beyond the bounds of precise (or probabilistic) predictability and enters a chaotic mode. Thus, that time [31] can be calculated as

$\Delta t_{lyap} = 1/\lambda.$  (3)

If $\lambda \to 0$, implying that $\Delta t_{lyap} \to \infty$, then long-term accurate predictions are possible.

### 3. Data and Computation

3.1. Data and Gauging Locations

Daily streamflow values were obtained from the National Water Information System: Web Interface at [32], for the Brazos River in Texas (USA) with a drainage area of approximately 118,000 km$^2$, extending from eastern New Mexico to more than 1000 km southeast to the Gulf of Mexico [33]. Daily streamflow observations from 12 USGS stream gauges on the mainstream were obtained for a period from 1989 to 2016, when simultaneous data for all the stations were available. The geographical locations of gauging stations are depicted in Fig. 3, and basic statistics of data are given in Table 1.

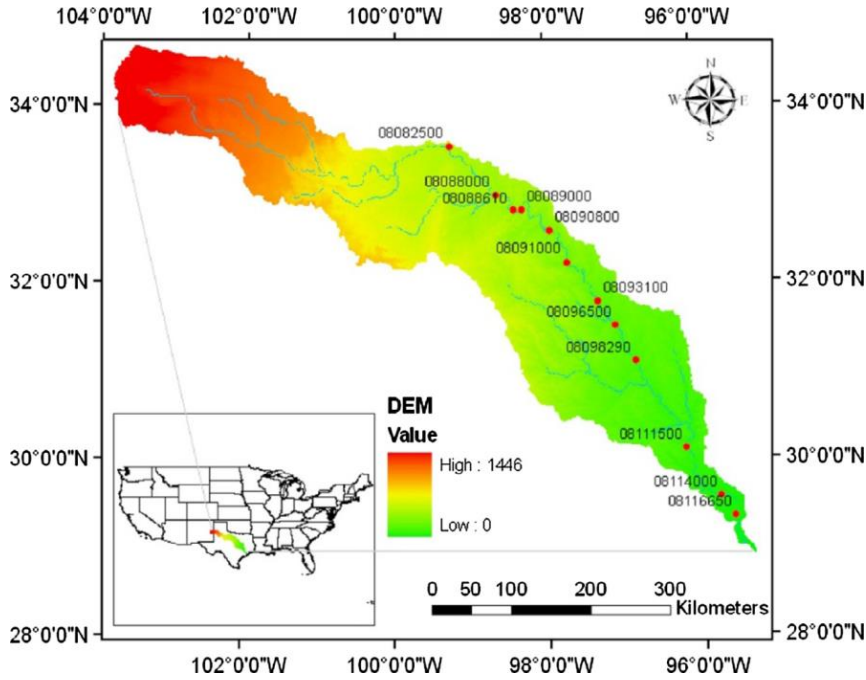

**Fig. 3.** Geographical locations of gauging stations on Brazos River used in this study [32].

**Table 1.** Basic descriptive statistics of daily discharge data of Brazos River for the period (1989-2016); (The first number indicates the order of the station used in this study).

| USGS Code | Station | Mean | Median | Min | Max | IQR | $SD_I$ |
|---|---|---|---|---|---|---|---|
| 1_08082500 | Seymour | 223.5 | 51.0 | 0.0 | 30700.0 | 130.0 | 907.9 |
| 2_08088000 | South Bend | 613.0 | 110.0 | 0.0 | 43800.0 | 320.0 | 2209.8 |
| 3_08088610 | Graford | 623.5 | 109.0 | 4.1 | 43800.0 | 300.0 | 2306.9 |
| 4_08089000 | Palo Pinto | 723.7 | 133.0 | 8.5 | 39700.0 | 361.0 | 2557.9 |
| 5_08090800 | Dennis | 974.4 | 195.0 | 0.0 | 79500.0 | 418.0 | 3600.3 |
| 6_08091000 | Glen Rose | 1078.8 | 86.0 | 1.5 | 82100.0 | 530.0 | 4093.9 |
| 7_08093100 | Aquilla | 1561.2 | 445.0 | 1.2 | 27100.0 | 1118.0 | 3687.3 |
| 8_08096500 | Waco | 2456.1 | 695.0 | 0.5 | 44000.0 | 1775.0 | 5237.7 |
| 9_08098290 | Highbank | 3103.7 | 873.5 | 30.0 | 70300.0 | 2240.0 | 6148.1 |
| 10_08111500 | Hempstead | 8014.3 | 2520.0 | 58.0 | 137000.0 | 7650.0 | 12821.1 |
| 11_08114000 | Richmond | 8523.8 | 2855.0 | 182.0 | 102000.0 | 8660.0 | 13232.0 |
| 12_08116650 | Rosharon | 8851.4 | 3060.0 | 27.0 | 109000.0 | 9080.0 | 13638.0 |

Daily discharge data were standardized, i.e., for each calendar day "$i$" mean discharge $\langle x_i \rangle$ and standard deviation $SD_i$, over the year, "$j$", were computed and then the standardized discharge on day "$i$" in year "$j$" was calculated as $y_{i,j} = (x_{i,j} - \langle x_i \rangle)/SD_i$ [9]. This procedure removes any seasonal effects.

3.2. Basic Descriptive Statistics

Basic descriptive statistics of daily discharge data of the gauging stations are summarized in Table 1, where for each station mean, median, minimum, maximum, interquartile range (IQR), and standard deviation ($SD_i$) are shown. It is seen from Table 1 that the differences between the maximum and the mean are in the range of roughly 10 to 40 standard deviations, strongly positively skewed, indicating a power law behavior. Indeed, frequency counts for the USGS 08082500 Brazos River station at Seymour, TX (USA), displayed in Fig.4 on a log-log scale demonstrate a power law distribution, with similar behavior also observed at all other stations.

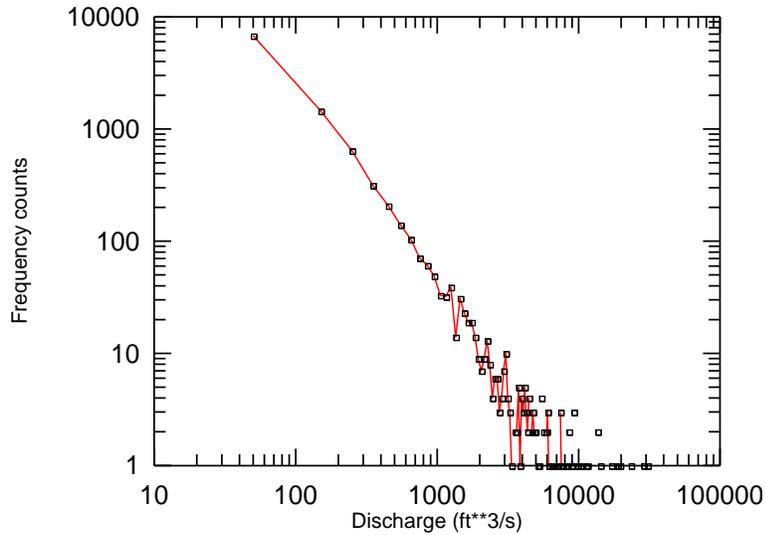

**Fig. 4.** Frequency counts of daily discharge data for the USGS 08082500 Brazos River station at Seymour, TX (USA) for the period (1989-2016).

## 4. Results and Discussion

### 4.1. Features of Streamflow Time Series

Before analyzing streamflow complexity, flow characteristics were considered to establish potential dissimilarities between different streamflow series. The box and whisker plot is useful for comparing distributions of datasets [34]. Figure 5 depicts a box and whisker plot of standardized daily discharge data and gives a clearer picture of the distribution of data in comparison with Table 1. It is seen that there is a smaller interval of variation at the third site, as well as a higher frequency of values smaller than the first quartile. The distributions of data for all stations have heavy right tails, stemming from a power law behavior.

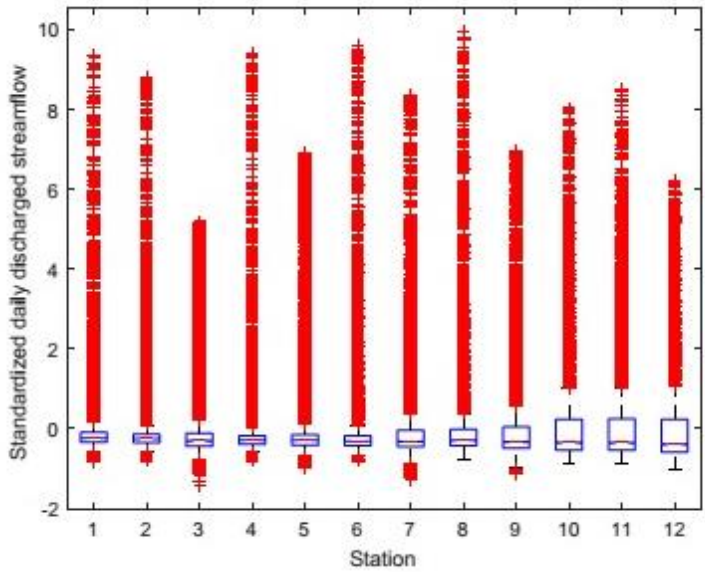

**Fig. 5.** Box and whisker plot of standardized daily discharge data on Brazos River.

All streamflow time series were characterized with a slow decay of positive autocorrelation. The values of lag $k$ such that $|r_k| < 1/e$ presented in Table 2 indicate that the rate of autocorrelation decline was lower for sites (7-12).

**Table 2.** Values of lag $k$, exponent α and Hurst exponent $H$ for standardized daily discharge data on Brazos River

| USGS Code | $k$ | $\alpha$ | H | $H(R/S)$ |
|---|---|---|---|---|
| 1_08082500 | 13 | -0.48 | 0.76 | 0.79 |
| 2_08088000 | 9 | -0.47 | 0.76 | 0.81 |
| 3_08088610 | 10 | -0.40 | 0.80 | 0.87 |
| 4_08089000 | 7 | -0.48 | 0.76 | 0.81 |
| 5_08090800 | 9 | -0.49 | 0.76 | 0.82 |
| 6_08091000 | 8 | -0.48 | 0.76 | 0.81 |
| 7_08093100 | 32 | -0.41 | 0.79 | 0.85 |
| 8_08096500 | 32 | -0.43 | 0.79 | 0.84 |
| 9_08098290 | 31 | -0.40 | 0.80 | 0.84 |
| 10_08111500 | 42 | -0.36 | 0.82 | 0.83 |
| 11_08114000 | 43 | -0.36 | 0.82 | 0.83 |
| 12_08116650 | 52 | -0.33 | 0.84 | 0.84 |

.

If lag $k$ increased, autocorrelations followed a power law, i.e., $r_k \sim ck^\alpha$ if $k \to \infty$. Exponent $\alpha$ and Hurst exponent ($H$), as a measure of the long-term memory of a time series, were calculated using the function *hurstACVF(x)* of the R package *fractal* version 2.0-1 [35], from statistical program R 3.4.2. The method for estimating $H$ was based on linear regression of scaled $asinh$ of autocovariance function (ACVF) versus $log(lag)$ over intermediate lag values. The values of $H$, that were between 0.5 and 1 (Table 2), indicated that the streamflow time series contained long memory or long range dependence [36]. This conclusion confirmed Hurst $H(R/S)$ obtained by means of Rescaled Range Method using the function *hurstex(x)* of the *R* package *pracma*. Analysis showed that the standardized daily discharge time series exhibited a long memory with the tendency to increase with the order of station except for site 3_08088610 where the series had a longer memory than did most stations. It should be noted that long range dependence may be spurious and be a consequence of nonstationarity, nonlinearity, and structural changes of streamflow [37]. Unlike random noise; long memory allows at least short-term predictability. Thus, one cannot conclude randomness of streamflow at individual stations.

## 4.2. Kolmogorov Complexity Measures and Kolmogorov Complexity Spectrum of Standardized Discharge Daily Data

The Kolmogorov complexity measures applied in this work shed additional light on the complex behavior of streamflow, complementing the usual entropy approaches commonly found in the literature. The values of KC, KCM, and MKC of standardized daily discharge data are shown in Fig. 6 which shows that the KC values for all daily streamflow time series were relatively small, ranging in the interval (0.20; 0.47). A similar behavior was observed for KCM and MKC, having values in the intervals (0.25; 0.68) and (0.08; 0.29), respectively. This is expected for lowland rivers in contrast to mountain rivers whose KC values can be up to 0.98 [8]. This feature of lowland rivers is clearly visualized through MKC (blue squares),

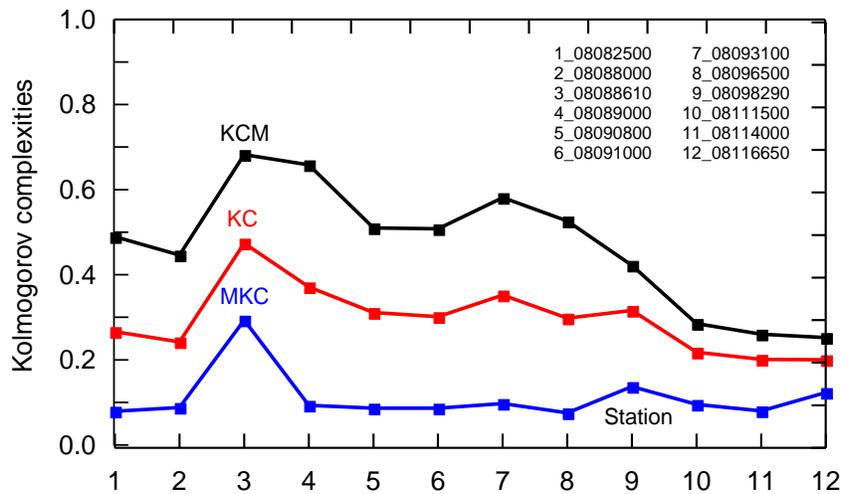

**Fig. 6.** Kolmogorov complexity (KC), the highest value of Kolmogorov complexity spectrum (KCM), and mean Kolmogorov complexity (MKC) of standardized daily discharge data on Brazos River.

which, on average, did not exceed 0.15. Although MKC offers some insight into the streamflow complexity, it is still not sufficiently reliable because it is obtained as an average value from the Kolmogorov spectrum, indicating that some details have been lost. However, by KC (red squares) and KCM (black squares) more content information measures can be obtained. Between these two information measures, KC yields less information about streamflow complexity than does KCM. That is, KC gives average information about the streamflow time series complexity, since its complexity remains hidden in the rules of the applied procedure. In contrast, KCM contains information about the highest complexity among all complexities observed in the KC spectrum, indicating the highest level of streamflow randomness. Thus, it can be said that MKC,

KC and KCM are overall, average and highest measures, respectively, of the complexity of streamflow time series. Finally, Fig. 6 shows that all three measures have the highest value, indicating that the streamflow time series at site 3_08088610 has the highest complexity.

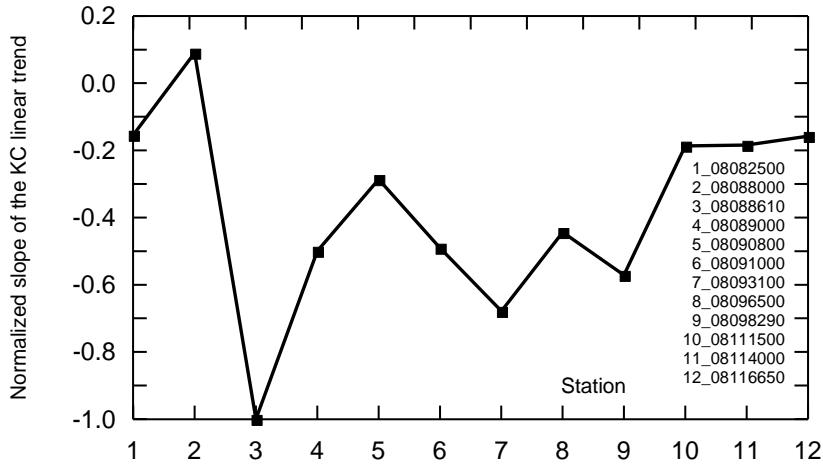

**Fig. 7.** Normalized slope of the KC linear trend of standardized daily discharge data of Brazos River.

The above measures can be used to estimate the behavior of a long streamflow time series for (1989-2016, in this study). For example, a decreasing KC is a reliable indicator that the streamflow series becomes more uniform due to either environmental or human influences. First, RKC was applied to calculate the KC of daily streamflow data for all sites. Then, the slope of linear trend was obtained for each site and then all slopes were normalized by the highest absolute value. These values are depicted in Fig. 7 which shows that: (i) the normalized slopes of KC linear trends of streamflow time series for all the sites were negative (ranged between -1 and -0.15), except for site 2_0808800 whose normalized slope of the KC linear trend was positive (0.09); and (ii) site 3_08088610 had the highest negative value of the KC linear trend (-1). Apparently, there exists a systematic decrease of the KC along the Brazos River, except for site 2_0808800, likely due to the aforementioned reasons.

Figure 8 shows the Kolmogorov complexity spectra of standardized daily discharge data. Inspection of Figs. 8a-8b shows that the site with the highest KCM corresponded to the shortest range (Table 1) (site 3_08088610). In contrast, sites 4_8089000 and 7_08093100, also having high values of the KCM, covered a wider range. Quite naturally, a question arises: What does this spectrum offer? It gives information about the randomness of each amplitude in the streamflow time series using all information, i.e. all samples representing the time series.

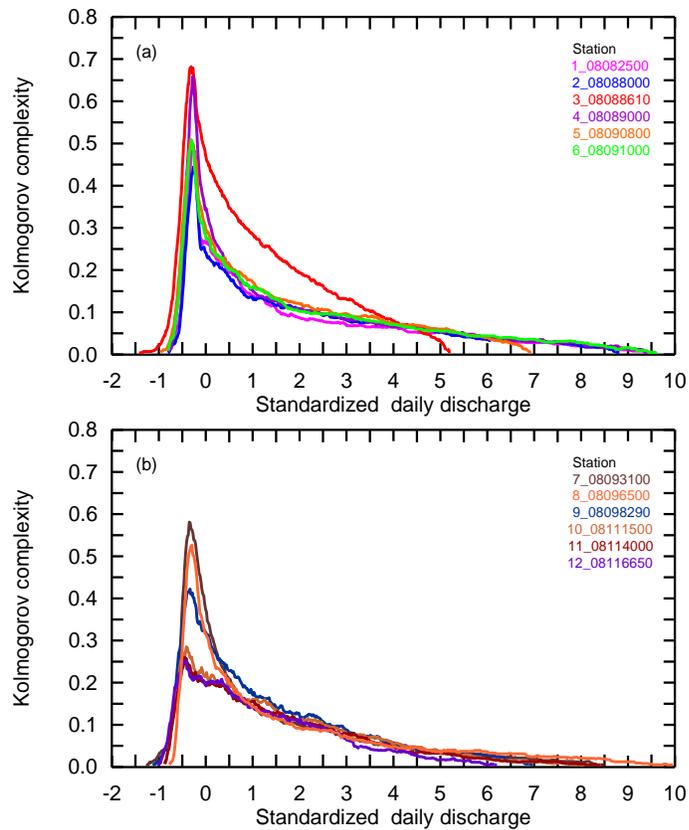

**Fig. 8.** Kolmogorov complexity spectra of standardized daily discharge data of Brazos River.

Figure 9 can be understood as a "streamflow discrete spectrum of the highest complexity" of standardized daily discharge data. This spectrum describes the presence of highest randomness (no other randomness left) in streamflow being influenced by many factors. This is a key element having important implications for modeling and prediction, since randomness is irreproducible and unpredictable [11,12].

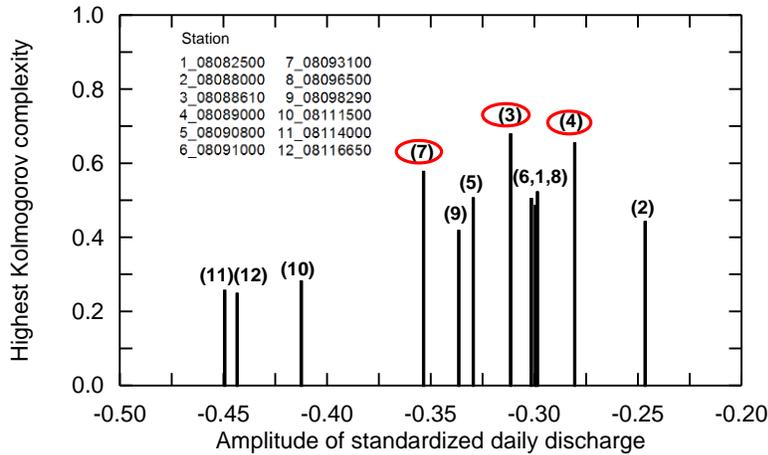

**Fig. 9.** The "streamflow discrete spectrum of the highest complexity" of standardized daily discharge data of Brazos River. The numbers in red ellipses correspond to the KC complexity spectrum highest values for three sites.

In the above analysis we mentioned that change of the KC complexity can be significantly induced by human activities. In that sense, our attention will be focused on the streamflow time series for site 3_08088610 since it has the highest values for all three KC complexity measures. There are many factors, both natural (runoff from rainfall and snowmelt; evaporation from soil and surface-water bodies; transpiration by vegetation; ground-water discharge from aquifers; ground-water recharge from surface-water bodies; sedimentation of lakes and wetlands; and formation or dissipation of glaciers, snowfields, and permafrost) and human-induced (surface-water withdrawals and transbasin diversions; river-flow regulation for hydropower and navigation; construction, removal, and sedimentation of reservoirs and stormwater detention ponds; stream channelization and levee construction; drainage or restoration of wetlands; land-use changes such as urbanization that alter rates of erosion, infiltration, overland flow, or evapotranspiration; wastewater outfalls; and irrigation wastewater return flow), that cause continuous changes in streamflow time series and therefore in its complexity. All of the above factors can contribute to randomness either singly or in synergy with other factors. The reason for the high KC complexity of this station may be attributed to the Morris Sheppard Hydroelectric Power plant at Morris Sheppard Dam (Possum Kingdom Reservoir) on the Brazos River in Palo Pinto County, built in the period 1938-1941 (11 miles southwest of Graford and 18 miles northeast from Mineral Wells). Currently, "USGS station 08088610 (Brazos River near Graford, TX) is located approximately 1.25 miles downstream of Possum Kingdom Reservoir. As such, this site is largely influenced by regulation. This gauge was established to monitor outflow from Possum Kingdom Reservoir. The gauge was initially located farther upstream, closer to the outflow from the reservoir. In 1995, the gage was moved downstream to the current location" (information obtained from [32]). Note that [8], using the KC complexity analysis, found that during 1946–1965, there was a change in complexity in the River Miljacka and the River Bosnia in comparison to the other chosen subintervals (analyzed time series was for the period 1926-1990). That complexity loss was interpreted as the result of intensive different human interventions on those rivers (establishing the network of channels and small dams for building the capacities for water consumption) after the Second World War.

4.3. Lyapunov Exponent and Predictability of Standardized Daily Discharge Data

Figure 10 plots the Lyapunov exponent of standardized daily discharge data and shows that station 3_08088610 has the highest value of LE (0.3937), while all other sites have values in the interval (0.0183; 0.1585). In the last decade, there has been an increasing use of LE in daily streamflow analysis for detecting the deterministic chaos [13,14] and improving the chaotic approach to accurately predict potential future outgrowths. Using LE as an indicator, [13] investigated the potential existence of chaos in daily streamflow time series of Kizilirmak River (Iran). They established the presence of low chaos in the flow series with a positive value of LE (0.0061). Considering forecasting streamflow of Xijiang River (China) based on chaos radial basis function networks, [15] reported that LE was 0.1604. For all Brazos River sites, the LE values can be placed between the LE values of Kizilirmak River and Xijiang River. Only exception is site 3_08088610 having a much higher value. This feature as well as a longer memory than the most stations can be attributed to structural changes in streamflow resulting from human and environmental activities.

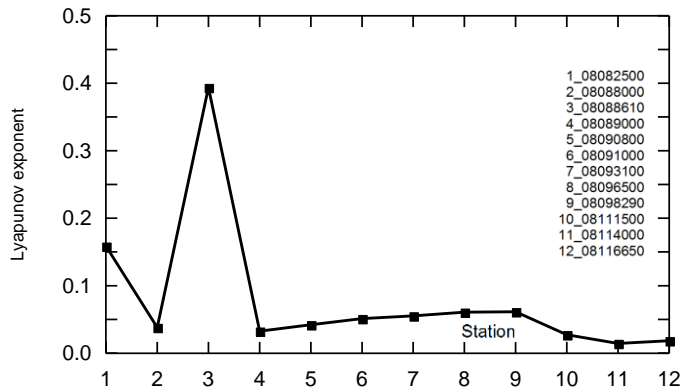

**Fig. 10.** Lyapunov exponent of standardized daily discharge data of Brazos River.

Using the LE method, one can obtain the Lyapunov time, $\Delta t_{lyap}$ (LT) i.e. the time interval for making accurate predictions [15,38-40, among others]. Figure 11a depicts the LE method for 12 sites on Brazos River given by the LT in days having the shape of a power function. Simple inspection indicates that the lowest predictability was for sites 3_08088610 (3 days) and 1_08082500 (6 days). Sites 8_08096500, 9_08098290, 7_08093100 and 6_08091000 had predictability of less than 20 days, while for sites 5_08090800, 2_08088000, 4_08089000 and 10_08114000, the prediction effect was between 20 and 40 days. Finally, the lowest LE values were for sites 11_08114000 and 12_08116650. Using this method [15] found that the prediction effect for the Xijiang River was poor (five-day prediction).

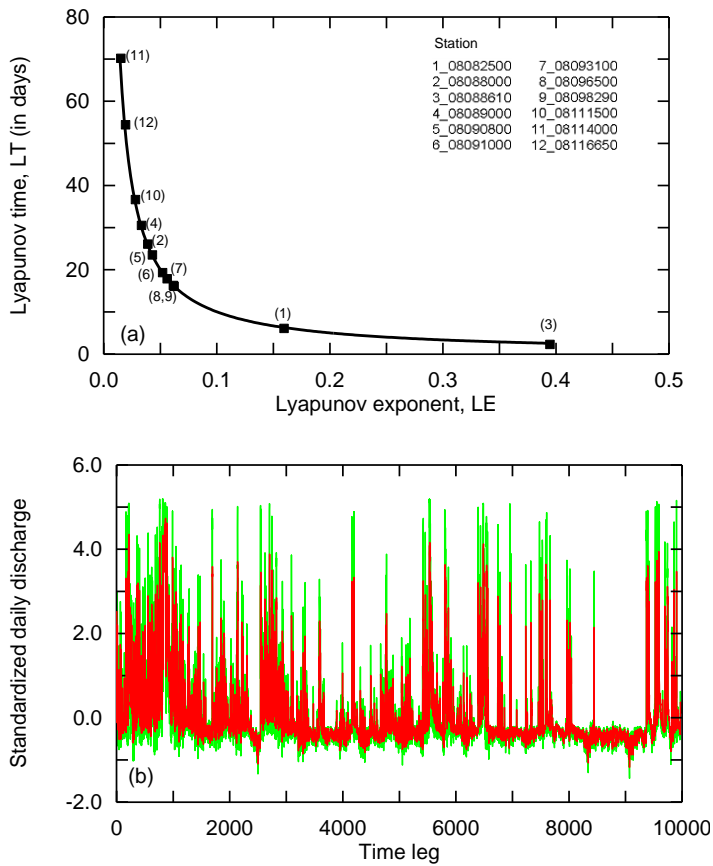

**Fig. 11.** (a) Predictability of standardized daily discharge data of Brazos River given by the LT (in days); (b) Standardized discharge daily streamflow for the site 3_08088610: measured (green) and simulated by the ARIMA model [41].

To illustrate the meaning and importance of the LT, which is not a predictability tool we applied an autoregressive integrated moving average (ARIMA (p, d, q)) model in its improved version [41] which is widely applied in hydrological simulations. This model has the property that uncertainty in a forecast increases exponentially with time. For the streamflow time series at site 3_08088610 the best ARIMA (5,1,3) model according to information criterion is identified, estimated and used for prediction h-steps ahead (h =1 in our simulation). The results of simulations are given in Fig. 11b. From this figure is seen that the fact that this model does not include the property of non-linearity which is a deterministic characteristic as well as a variable variance and the appearance of extreme values, leads to erroneous prediction. Note that LT for the site 3_08088610 indicates that already after three days any conceivable model correctly representing the phenomenon should enter a chaotic mode and cannot capture the non-linearity of the process. The smallest predictability in the case of the site 3_08088610 is not contradictory with long memory that characterizes this station. Long memory may be present in stochastic and deterministic processes. The ARFIMA model is an example of stochastic long memory process. In that case the maximal Lyapunov coefficient is zero, and the *H* value may be high. On the other hand, in some special cases of chaotic time series, there is positive correlation between max LE and *H* [42]. *H* and max LE are complementary measures, as *H* cannot measure sensitivity to initial conditions. So the long memory does not imply high predictability if the model is sensitive to initial conditions.

The Lyapunov time was corrected for the presence of randomness, since the chaos theory generally deals with "irregular behavior in a complex system that is generated by nonlinear deterministic interactions with only a few degrees of freedom, where noise or intrinsic randomness does not play an important role" [43]. The Lyapunov time was used for predicting the time series. However, many streamflow time series are highly complex. Therefore, $\Delta t_{lyap}$ can be corrected for randomness in the following way. Similar to $\Delta t_{lyap}$ we can introduce a randomness time $\Delta t_{rand} = 1/K_c$ (in time units, second, hour or day), where $K_c$ is the Kolmogorov complexity. Henceforth, we shall denote this quantity Kolmogorov time (KT), as it quantifies the time span beyond which randomness significantly influences predictability. Then, the Lyapunov time corrected for randomness is defined as $[0, \Delta t_{lyap}] \cap [0, \Delta t_{rand}]$. It could be stated that the KT designates the size of the time window within time series where complexity remains nearly unchanged. Figure 12a shows the predictability of standardized daily discharge data given by the LT in days corrected for randomness. The figure shows that the prediction effect $\Delta t_{rand}$ for KC is extended between two and five days, while for KCM it is between one and four days.

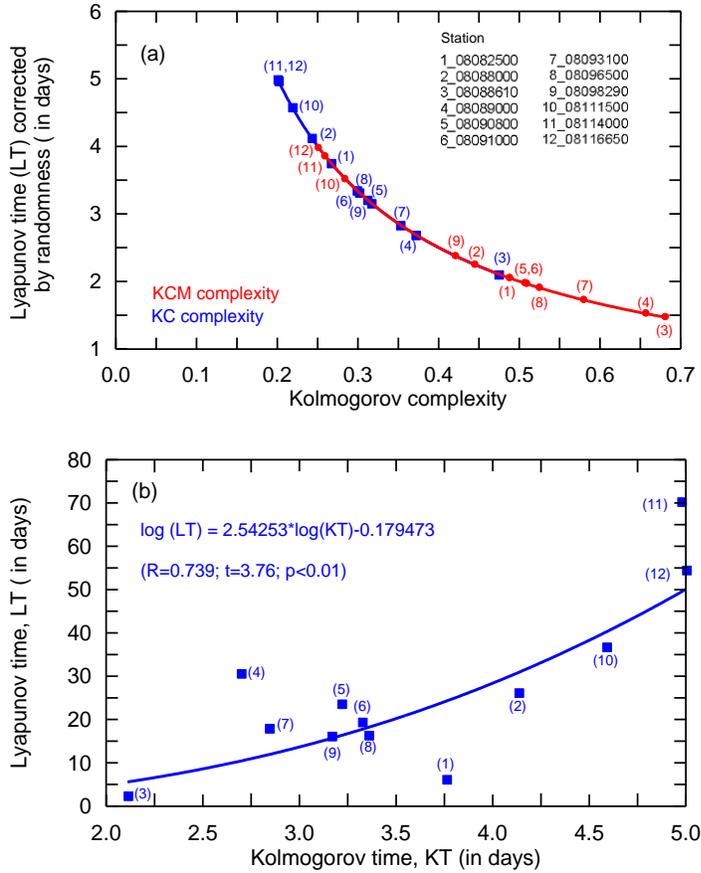

**Fig. 12.** (a) Predictability of standardized daily discharge data of Brazos River given by the Lyapunov time (LE) corrected by randomness (in days); (b) LT time versus KT time.

Figure 12b depicts Lyapunov time (LT) plotted versus Kolmogorov time (KT). It is seen that the relationship between them is expressed through a power law dependence. How much the randomness can reduce the LT can be seen from comparison of Figs. 11a and 12b. According to Fig. 11b the longest predictability in the LT units for the 12 sites at Brazos River is around 70 days, while following from the fit on Fig. 12b the presence of randomness reduces that number to 45 days.

## 5. Conclusions

Daily river flow data from twelve gauging stations on Brazos River in Texas (USA) is analyzed using the Kolmogorov complexity and related complexity measures (Kolmogorov complexity spectrum and its highest value, running, and mean Kolmogorov complexity), Lyapunov exponent and Lyapunov time. The following conclusions are drawn from this study:

(1) Standardized daily discharge time series exhibit long memory with a tendency to increase with the order of the gauging station with the exception of one station which has a longer memory than do most stations. Long range dependence may, however, be spurious and be the consequence of non-stationarity, nonlinearity and structural changes of time series. Unlike random noise, long memory allows for at least short-term predictability.

(2) The values of KC, KCM, and MKC of streamflow are relatively small, except for one site which has the highest value.

(3) All complexity measures can be used to estimate the behavior of streamflow time series for a long time period. Thus a decrease of KC is a reliable indicator that streamflow series becomes more uniform due to environmental or human influences. The normalized slope of KC linear trend of streamflow series for all sites is negative, except for a site whose normalized slope is positive. There exists a systematic decrease of KC along the Brazos River, except for one site.

(4) Kolmogorov complexity spectra yield information about the randomness for each amplitude in the streamflow time series.

(5) The streamflow discrete spectrum of the highest complexity shows the presence of highest randomness in river flow, a key in modeling and prediction.

(6) The Kolmogorov time (KT) designates the size of the time window within time series where complexity remains nearly unchanged.

**Acknowledgments**

This paper was realized as part of the project "Studying climate change and its influence on the environment: impacts, adaptation and mitigation" (43007) financed by the Ministry of Education and Science of the Republic of Serbia within the framework of integrated and interdisciplinary research for the period 2011–2018. We also acknowledge support of Brazilian agencies CAPES and CNPq (grant No 310441/2015-3).

**Funding**

This research did not receive any specific grant from funding agencies in the public, commercial, or not-for-profit sectors.